\documentclass[aps,pra,twocolumn,showpacs,floatfix]{revtex4}
\usepackage{psfrag}
\usepackage{graphicx}
\usepackage{amsbsy,latexsym}
\usepackage{amsfonts}
\usepackage{amssymb}
\usepackage[mathscr]{eucal}

\newcommand{\tr}{{\rm tr}\,}

\newcommand{\ket}[1]{\vert #1 \rangle}
\newcommand{\bra}[1]{\langle #1 \vert}

\newcommand{\D}[3]{\mathfrak D^{(#1)}_{#2 #3}}

\newcommand{\vn}{\vec{n}}
\newcommand{\vM}{\vec{M}}
\newcommand{\vMx}{\vec{M}(\chi)}
\newcommand{\vVx}{\vec{V}(\chi)}

\newcommand{\ketn}{\ket{\vec{n}}}
\newcommand{\bran}{\bra{\vec{n}}}
\newcommand{\kMx}{\ket{\vec{M}(\chi)}}

\newcommand{\NN}{{\mathscr N}}
\newcommand{\eqref}[1]{(\ref{#1})}
\begin{document}
\title{Estimation of  qubit  pure states
with collective and individual measurements}

\author{E.~Bagan, A.~Monras and R.~Mu{\~n}oz-Tapia}
\affiliation{Grup de F{\'\i}sica Te{\`o}rica \& IFAE, Facultat de
Ci{\`e}ncies, Edifici Cn, Universitat Aut{\`o}noma de Barcelona, 08193
Bellaterra (Barcelona) Spain}

\begin{abstract}
We analyze the estimation of a qubit pure state by means of local
measurements on $N$ identical copies  and compare its average
fidelity for an isotropic prior probability distribution to the
absolute upper bound given by collective measurements.  We discuss
two situations:  the first one,  where the state is restricted to
lie on the equator of the Bloch sphere, is formally equivalent to
phase estimation; the second one, where there is no constrain on
the state, can also be regarded as the estimation of a direction
in space using a quantum arrow made out of $N$ parallel spins. We
discuss various schemes with and without classical communication
and compare their efficiency.  We show that the fidelity of the
most general collective measurement can always be achieved
asymptotically with local measurements and no classical
communication.
\end{abstract}

\maketitle

\section{Introduction}\label{introduction}
In Quantum Information, measurement and estimation are not just
\emph{some} important topics. They are at the very core of the
theory. An unknown quantum state can only be unveiled by
means of measurements, and the information is always gained at the
expense of destroying the state. This information is then
processed to obtain the desired estimate.

Any estimation procedure requires a sample of identical copies of
the unknown state on which we can perform measurements. If unknown
states could be copied, then one could produce samples of an
arbitrary number of copies  and the state could be estimated with
infinite accuracy. The no-cloning theorem, however, prevents this
possibility~\cite{no-cloning}. But even in such unphysical
circumstances, one would only have finite  time and a limited
number of resources for copying and measuring. So, in the real
world only a finite, usually not large,  number of copies are
available and  only a reasonable approximation to the unknown
state can be made. It is thus very important to devise strategies
with optimal performance in a variety of practical situations.

Over the last few years, it has been recognized that a joint
measurement on $N$ copies is more efficient than $N$
individual measurements on each copy separately. We will often
refer to the former  as {\em collective}, in contrast  to {\em
individual} (also called {\em local})  measurements. The quantum
correlations behind the collective measurements are (almost)
always more powerful than the classical correlations used in
sequential individual measurements~\cite{pw,mp}. This and other
issues have been studied in various
contexts~\cite{holevo,helstrom,book,braunstein,derka,
lpt,direction-1,bbm-direction,ps-direction,
ajv,bbm-reference-1,bbm-reference-2,ps-reference,bartlett}, but
for local measurements not many analytical results have been
obtained~\cite{jones,gill-massar,fkf,hannemann,bbm-local,bbm-mixed,
embacher}. The most powerful, involve sophisticated estimation
theory technology that is not so widely known among physicists.
Moreover, they mainly apply in the asymptotic limit (when $N$ is
large), and in a ``pointwise" fashion, in the sense that no average
over the prior probability distribution of unknown states is
considered~\cite{japos-1,japos-2}.

Here we address the issue of estimating the most elementary
quantum pure state, that of a  qubit, assuming we have a sample of
$N$ identical copies of it. We consider two relevant cases:
estimation of a completely unknown qubit state, i.e. one given by
an arbitrary point on the Bloch sphere; and estimation of
restricted states that are known to lie on the equator of the
Bloch sphere. The latter is also interesting because it is
formally equivalent to phase estimation (these states are also
equivalent to the so-called rebits~\cite{rebits}). We refer to
these two situations as 3D and 2D cases, respectively.

The most general measurement is described by a Positive Operator
Valued Measure (POVM) on the $N$ copies.  An optimal measurement
of this type yields the ultimate bounds that can be achieved by
any estimation procedure. We will re-derive these bounds in a
unified, very comprehensive framework. For local measurements, we
will consider von Neumann measurements, as these can be readily
implemented in a laboratory. So, by ``local measurement"  we will
loosely mean a \emph{local von Neumann measurement}.  Furthermore,
we will show that
for some of the local procedures discussed here, optimal
measurements are necessarily of von Neumann type. We will refer to
local schemes that use classical communication, i.e. those that
exploit the possibility of  using the actual outcomes that are being obtained
to dynamically adapt the next measurements, as LOCC (Local Operations and
Classical Communication) schemes.

Let us be more concrete about the problem we are concerned with.
Assume that we are given an ensemble of $N$ identical copies of an
unknown qubit state, which we denote by $\ketn$, where $\vn$ is
the  unit vector on the Bloch sphere that satisfies
\begin{equation}\label{bloch}
    \ketn\bran=\frac{1+\vn \cdot \vec{\sigma}}{2},
\end{equation}
and $\vec{\sigma}=(\sigma_x,\sigma_y,\sigma_z)$ are the usual
Pauli matrices. After performing a measurement (collective or
local) on the $N$ copies of $\ketn$, one obtains an
outcome~$\chi$. Based on~$\chi$, an estimate, $\kMx$, for the
unknown state is guessed. To quantify how well $\kMx$ approximates
the unknown state $\ketn$ we use the fidelity, defined as the
overlap
\begin{equation}\label{f}
  f_n(\chi)\equiv|\bra{\vn}\vMx \rangle|^2={1+\vn \cdot\vMx \over
  2}.
\end{equation}
The fidelity~\eqref{f} can be regarded as a ``score": we get ``1"
for a perfect determination ($\vec{M}=\vec{n}$) and ``0" for a
completely wrong guess ($\vec{M}=-\vec{n}$). Our aim is to
maximize the average fidelity, hereafter fidelity in short, over
the initial probability and all possible outcomes,
\begin{equation}\label{f-1}
    F\equiv \langle f \rangle =
\sum_\chi \int dn\, f_n(\chi)\;p_{n}(\chi),
\end{equation}
where $dn$ is the prior probability distribution, and
$p_{n}(\chi)$ is the probability of getting  outcome $\chi$ given
that the unknown state is~$\ket{\vec{n}}$.

As mentioned above, we allow for general measurements or POVMs.
They are defined by a set of positive operators $\{O(\chi)\}$
(each one of them associated to an outcome) that satisfy the
condition
\begin{equation}\label{povm}
   \sum_\chi O(\chi)=\openone.
\end{equation}
The probability $p_{n}(\chi)$ is given in terms of these operators
by
\begin{equation}\label{pnx}
    p_{n}(\chi)=\tr[\rho_n O(\chi)],
\end{equation}
where $\rho_n$ is the quantum state of the $N$ identical copies,
i.e., $\rho_n=(\ketn \bran)^{\otimes N}$.

 In Eq.~\eqref{f-1} there are only two elements that require optimization: the guess
and the POVM, which enter this equation through~(\ref{f})
and~(\ref{pnx}), respectively. The optimal guess (OG) can be obtained
rather trivially. The Schwarz inequality shows that the choice
\begin{equation}\label{m-optimal}
    \vMx=\frac{\vVx}{|\vVx|},
\end{equation}
where
\begin{equation}\label{v-optimal}
    \vec{V}(\chi)=\int dn\; \vec{n}\; p_n(\chi),
\end{equation}
maximizes the value of the fidelity, which then
reads~\cite{bbm-local}
\begin{equation}\label{f-optimal}
    F=\frac{1}{2}\left(1+\Delta
    \right)\equiv\frac{1}{2}\left(1+\sum_\chi|\vVx|\right)
\end{equation}
(in the strict sense we should write $F_{\rm OG}$, but  to
simplify the notation, we drop the subscript when no confusion
arises). Eq.~\eqref{m-optimal} gives the best state that can
be~inferred, and Eq.~\eqref{f-optimal} gives the maximum fidelity
that can be achieved for \emph{any} prior probability and
\emph{any} measurement scheme specified by the conditional
probabilities $p_{n}(\chi)$. We are thus left only with the
non-trivial task of obtaining the optimal measurement. The goal of
this paper is to compute the maximum value of~\eqref{f-optimal}
within a unified framework for various measurement schemes,
specially the local ones.

The paper is organised as follows. In the next section we derive
the bounds on the fidelity of the optimal collective measurements.
In Sec.~\ref{local} we
discuss several local measurement schemes, with and without
classical communication. Asymptotic values of the fidelity  are
computed in Sec.~\ref{asymptotic}. A summary of results and our
main conclusions are presented in Sec.~\ref{conclusions}, and two
technical appendices end the paper.

\section{Collective measurements}\label{collective}
\subsection{2D states}\label{2D states}
A 2D state corresponds to a point that is known to lay on the
equator of the Bloch sphere.  If we take it to be on the $xy$
plane, such state, $\ketn$, has $\vn=(\cos\theta,\sin\theta,0)$.
If no other information is available, the prior probability distribution has
to be isotropic, that is, $dn=d\theta/(2\pi)$.
Eq.~\eqref{f-optimal} then reads
\begin{equation}\label{Delta-2D}
   \Delta=\sum_\chi\left|\int \frac{d\theta}{2 \pi}\vn \ \tr[\rho_n O(\chi)]\right|.
\end{equation}
Notice that we can write,
\begin{equation}\label{rhon}
    \rho_n\equiv\rho(\theta)=U(\theta)\rho_0 U^{\dag}(\theta),
\end{equation}
where $\rho_0$ is a fiducial state of angular momentum $J\equiv N/2$ and
maximal magnetic  quantum number, $m=J$, along any fixed direction
on the equator of the Bloch sphere. In particular,  $\rho_0$ can
be chosen to point along the $x$ axis,
\begin{equation}\label{rhoxn}
   \rho_0={\ket{JJ}_x}\,  {}_x\bra{JJ}.
\end{equation}
In Eq.~\eqref{rhon}, $U(\theta)$ is the unitary representation of a
rotation around the $z$ axis. The group of such unitary matrices is
isomorphic to $U(1)$.  In the standard basis $\ket{jm}\equiv \ket{m}$,
where $U(\theta)$ is diagonal, we have
\begin{equation}\label{rho-theta}
    \rho(\theta)=\sum_{m\ n}e^{i (m-n)\theta}\left({\rho_0}\right)_{m n}
    \ket{m}\bra{n}.
\end{equation}
We are now ready to compute \eqref{Delta-2D}:
\begin{eqnarray}\label{Delta2D-1}
   \Delta&=&\sum_\chi\left|\sum_{m n} \int \frac{d\theta}{2 \pi} e^{i \theta} e^{i
   (m-n)\theta}
   \left({\rho_0}\right)_{m n} [O(\chi)]_{n m} \right| \nonumber \\
    &=&\sum_\chi\left|\sum_{m=-J}^{J-1}
   \left({\rho_0}\right)_{m m+1} [O(\chi)]_{m+1\, m} \right|,
\end{eqnarray}
where $[O(\chi)]_{mn}\equiv\bra{m}O(\chi)\ket{n}$. The following
inequalities give the maximum value of $\Delta$:
\begin{eqnarray}
    \nonumber
   \Delta &\leq&  \sum_\chi\sum_{m=-J}^{J-1}
   \left|\left({\rho_0}\right)_{m m+1} [O(\chi)]_{m+1\, m} \right|\\
   \nonumber
   &\leq& \sum_{m=-J}^{J-1}
   \left|\left({\rho_0}\right)_{m m+1}\right| \sum_\chi\left|[O(\chi)]_{m+1 \,m} \right|\\
   &\leq& \sum_{m=-J}^{J-1}
   \left|\left({\rho_0}\right)_{m m+1}\right|,
   \label{optimal-Delta-2D}
\end{eqnarray}
where in the last step in~(\ref{optimal-Delta-2D}) we have used
that
\begin{equation}\label{povm-cond}
   \sum_\chi\left|[O(\chi)]_{m+1 m} \right|\leq 1,
\end{equation}
as follows from positivity  and~(\ref{povm}). More precisely,
positivity implies
\begin{equation}\label{positivity}
    [O(\chi)]_{m m}[O(\chi)]_{m+1\, m+1}\geq \left|[O(\chi)]_{m
    m+1}\right|^2,
\end{equation}
and  the Schwarz inequality yields
\begin{eqnarray}
    &\displaystyle \sum_\chi \left|[O(\chi)]_{m \,m+1}\right|\leq &\nonumber \\
   &\displaystyle  \sqrt{\sum_\chi
    [O(\chi)]_{m m}}\sqrt{\sum_\chi[O(\chi)]_{m+1\, m+1}}=1 .&
    \label{schwarz-povm}
\end{eqnarray}

There are two points worth emphasizing here. First, all the
inequalities can be saturated, therefore the bound is tight (we
give below a POVM that accomplishes this task). Second, the bound
\eqref{optimal-Delta-2D} is completely general. If we were
interested in encoding the information carried by a phase $\theta$
in a covariant way [as in \eqref{rhon}], but using a general
fiducial state~\cite{braunstein,wiseman,wiseman-2},
$\rho_0=\sum_{mm'}a_m a^*_{m'}\ket{m}\bra{m'}$ (ideally the best
possible one, which is not necessarily a tensor product of
identical copies), the fidelity would still be bounded by
\eqref{optimal-Delta-2D}. In this general case, one has
$\Delta\le\sum_{mm'}a^*_{m'}\mathsf{M}_{m'm}a_m$, where
$2\mathsf{M}_{m'm}=\delta_{m'\,m+1}+\delta_{m'+1\,m}$, and its maximum
value  is given by the largest eigenvalue of the matrix $\mathsf M$. A
straightforward calculation gives~\cite{wiseman}
$F_{\max}=(1+\cos[2\pi/(d_J+2)])/2$, where $d_j=2j+1$ is the
dimension of the invariant Hilbert space corresponding to the
representation ${\bf j}$ of $SU(2)$.

In our problem $\rho_0$ is constrained by the condition of having
identical copies. From \eqref{rhoxn} and \eqref{optimal-Delta-2D}
one has~\cite{derka}
\begin{eqnarray}\label{Delta-2D-max}
    \Delta& \leq &\frac{1}{2^{N}}\sum_{m=-J}^{J-1}
    \sqrt{\pmatrix{N \cr J+m}\pmatrix{N \cr J+m+1}}\nonumber\\
   &=& \frac{1}{2^N}\sum_{m=-J}^{J} \pmatrix{N \cr
   J+m}\sqrt{\frac{J-m}{J+m+1}} ,
\end{eqnarray}
where we have used that
\begin{equation}
\ket{JJ}_x=\ket{\vec x}^{\otimes N}={1\over
2^J}\sum_{m=-J}^J\pmatrix{N\cr J+m}\ket{Jm}
\end{equation}
(recall that $J\equiv N/2$).

 We next show that there are POVMs that
attain this bound. To saturate the first inequality
in~(\ref{optimal-Delta-2D}) the phase of $[O(\chi)]_{m\,m+1}$ must
be independent of $m$. This is ensured if this phase
is a function of $m-n$. Similarly, a set of
positive operators for which \mbox{$|O(\chi)_{m
n}|=\mathit{constant}$} for all $\chi$, $m$ and $n$, will
certainly saturate the remaining inequalities
in~(\ref{optimal-Delta-2D}). In particular, the covariant
(continuous) POVM, whose elements are given by
\begin{equation}
[O(\phi)]_{mn}={\rm e}^{i(m-n)\phi} , \label{optimal-povm-2D}
\end{equation}
satisfies all the requirements. Note that we have labeled the
outcomes by a rotation angle~$\phi$, which plays the role of
$\chi$. Hence, condition~(\ref{povm}) becomes
\begin{equation}
 \int \frac{d\phi}{2 \pi} O(\phi) =\openone,
\label{optimal-povm-2D-old}
\end{equation}
which certainly holds for~(\ref{optimal-povm-2D}). These are rank
one operators, and can also be written as
\begin{equation}
O(\phi)=U(\phi)\ket{B}\bra{B}U^\dagger(\phi),
\end{equation}
where
\begin{equation}\label{B}
   \ket{B}=\sum_{m=-J}^{J}\ket{J,m}.
\end{equation}

We have just  shown that at least one optimal measurement
[i.e., a POVM that saturates~(\ref{optimal-Delta-2D})] exists, but
other optimal measurements can be found. POVMs with  finite number
of outcomes, for instance, are straightforward to obtain by
choosing $\phi$ to be the $d_J$-th roots of unity, namely,
\begin{equation}
[O(k)]_{mn}={1\over d_j}\exp\left\{{i(m-n){2\pi k\over
d_J}}\right\} ,
\end{equation}
where $k=1,\dots, d_J$. In this case $\{O(k)\}$ is a von
Neumann measurement, since the number of rank one POVM elements
equals the dimension of the Hilbert space.

In the asymptotic limit the fidelity can be obtained in terms of
the moments of a binomial distribution $ \mathrm{Bin}(n,p)$
with parameters $n=N$ and $p=1/2$. We simply need to
expand~\eqref{Delta-2D-max} in powers of $m$, i.e., around $\langle
m\rangle=0$, to obtain
\begin{eqnarray}
   \Delta&\leq&\frac{1}{2^N}\sum_m \pmatrix{N \cr
   J+m}\times  \\
   &&\left[1- \frac{2 m}{N}+\left( \frac{2
   m^2}{N^2}-\frac{1}{N}\right)+ O(1/N^{3/2})\right]\nonumber
   \label{Delta-2D-expansion}
\end{eqnarray}
(notice that sum over $m$ is shifted by $J$ with respect to the
usual binomial distribution). The  moments are well known to be
$\langle 1 \rangle=1$, $\langle m \rangle=0$ and \mbox{$\langle
m^2 \rangle=N/4$}. The latter shows that $m$ has ``dimensions" of
$\sqrt{N}$, which helps to organise the expansion in powers of
$1/N$. We finally obtain
\begin{equation}\label{Delta-2D-asymptotic}
    \Delta_{\max}=1-\frac{1}{2 N}+\cdots,
\end{equation}
and a fidelity
\begin{equation}\label{fidelity-2D-asymptotics}
    F=1-\frac{1}{4 N}+\cdots.
\end{equation}

\subsection{3D states}\label{3D states}
A 3D state $\ket{\vec n}$ corresponds to a general point on the
Bloch sphere. As in the previous section we write
\begin{equation}\label{rhon-3D}
    \rho_n\equiv\rho(\vec{n})=U(\vec{n})\rho_0 U^{\dag}(\vec{n}),
\end{equation}
where for convenience $\rho_0$ is now chosen to point along the
$z$ axis, i.e.,
\begin{equation}\label{rhoz-3D}
   \rho_0=\ket{JJ}\bra{JJ},
\end{equation}
and $U(\vec{n})$ is the unitary representation [i.e., the element
of $SU(2)$] of the rotation
 that brings $\vec{z}$
into~$\vec{n}$ (a rotation around the vector $\vec z \times \vec
n$).

Recalling \eqref{pnx} and \eqref{v-optimal}, we have
\begin{equation}\label{v-3D}
    \vec{V}(\chi)=\int d n\,\vec{n}\; \tr[\rho_n O(\chi)],
\end{equation}
where $dn$ is the invariant measure on the two-sphere, e.g.,
\begin{equation}
\label{febc dn}
dn={d(\cos\theta)d\phi\over4\pi},
\end{equation}
where $\theta$ and $\phi$ are the
standard azimuthal and polar angles.
Notice that we can always define an operator $\Omega(\chi)$ in
such a way that
\begin{equation}\label{O-optimal-3D-1}
    O(\chi)=U[\vMx] \Omega(\chi) U^{\dag}[\vMx],
\end{equation}
where $\vMx$ is given by~\eqref{m-optimal}. Taking into account
that $\Delta$ is rotationally  invariant  one obtains
\begin{equation}\label{Delta-3D-1}
     \Delta=\sum_\chi\left|\int dn \; n_z \; \tr[\rho_n
   \Omega(\chi)\right|.
\end{equation}
We readily see  that  $n_z= \cos\theta=\D{1}{0}{0}(\vec{n})$,
where the rotation matrices $\D{j}{m}{m'}$ are defined in the
standard way, $\D{j}{m}{m'}(\vec n)=\bra{jm}U(\vec n)\ket{jm'}$.
We then  have
\begin{eqnarray}\label{Delta-3D-2}
  \Delta &=&\sum_\chi\left|\sum_{mm'}\int dn \;\D{1}{0}{0}(\vec{n})\;
  \left(\rho_n\right)_{mm'}\Omega_{m'
   m}(\chi)]\right|\nonumber \\
    &=& \sum_{m}\left[\sum_\chi \Omega_{mm}(\chi)\right]\int dn \;\D{1}{0}{0}(\vec{n})\;
    \left(\rho_n\right)_{m
    m},
\end{eqnarray}
where in the second equality we have used that
\begin{equation}\label{schur again}
\int
dn\,\D{1}{0}{0}(\vec{n})\left(\rho_n\right)_{mm'}=\delta_{mm'}\int
dn\,\D{1}{0}{0}(\vec{n})\left(\rho_n\right)_{mm},
\end{equation}
as follows from Schur's lemma after realizing that the left hand
side of~(\ref{schur again}) commutes with $U(\theta)$, the
unitary transformations defined right after~\eqref{rhoxn}. Recall that
these transformations, which are a $U(1)$
subgroup of $SU(2)$, have only one-dimensional irreducible
representations, labeled by the magnetic quantum number~$m$, thus
yielding relation~(\ref{schur again}).
In~(\ref{Delta-3D-2}) we have removed the absolute value as all
terms are positive (see below). Tracing \eqref{O-optimal-3D-1} one
obtains $\sum_\chi \tr\Omega(\chi)=\sum_\chi \tr O(\chi)=d_J$.
Therefore
\begin{eqnarray}
    \Delta&\leq& d_J \max_{m} \int dn \;\D{1}{0}{0}(\vec{n})\;
    \left(\rho_n\right)_{mm}\nonumber\\
   &=&\max_m\langle10;Jm|Jm\rangle^2
   ={J\over J+1},
   \label{Delta-3D-2p}
\end{eqnarray}
where in the second equality we have used that
$\rho_{mm}=\D{J}{m}{J}(\vec{n})\mathfrak{D}^{(J)*}_{m J}(\vec{n})$
and the well known orthogonality relations of the SU(2)
irreducible representations~\cite{edmonds}.
Recalling that $J\equiv N/2$  we finally get~\cite{mp}
\begin{equation}\label{fidelity-3D}
    F=\frac{N+1}{N+2},
\end{equation}
which for large $N$ behaves as
\begin{equation}\label{asymptotic-3D}
   F=1-\frac{1}{N}+\cdots \,.
\end{equation}

Let us finally give a POVM that saturates all the inequalities.
The maximum of the Clebsch-Gordan $\langle10;Jm|Jm\rangle$ in~(\ref{Delta-3D-2p}) occurs at $m=J$. Hence, to
attain the bound we need to choose
\begin{equation}\label{povm-3D}
\Omega_{mm}(\chi)=c_\chi \delta_{Jm},\quad \sum_\chi c_\chi=d_J,
\end{equation}
where the  coefficients $c_\chi$ are positive. This leads
straightforwardly to the optimal continuous POVM defined by
\begin{equation}\label{povm-3d-2}
   O(\vec{m})=d_J \,U(\vec{m})\ket{JJ}\bra{JJ}U^{\dag}(\vec{m})
\end{equation}
(one can check that $ \int dm\; O(\vec{m})=\openone$).
 POVM's with finite number
of elements can also be constructed. The only requirements
are~\eqref{povm-3D}  and, of course, \eqref{povm}. These
constraints translate into a series of conditions for the set of
directions $\{\vec{m}\}$, for which solutions with a finite number
of elements can be found. We refer the reader to the
literature~\cite{derka,lpt, bbm-direction,bbm-reference-1} for
details.

\section{Local measurements}\label{local}
Collective measurements, although very interesting from the
theoretical point of view, are difficult to implement in
practice. Far more interesting for experimentalists are individual
von Neumann measurements~\cite{hannemann,experiments}.

Individual von
Neumann measurements on qubits are represented by two projectors
\begin{equation}\label{vonneumann}
    O(\pm \vec{m})=\frac{1\pm \vec{m}\cdot \vec{\sigma}}{2},
\end{equation}
where $\vec{m}$ is a unit Bloch vector characterizing the
measurement (in a spin system, e.g., $\vec{m}$ is the orientation
of a Stern-Gerlach). In a general frame we must also allow
classical communication, i.e., the possibility of adapting the
orientation of the measuring devices depending on previous
outcomes~\cite{fkf,hannemann,bbm-local}.

In the next sections we study quantitatively several schemes in
ascending order of optimality: from the most basic tomography
inspired schemes~\cite{tomography,likelihood} to the most general
individual measurement procedure with classical
communication~\cite{bbm-local}.

Our aim is to investigate how good these local measurements are as
compared to the collective ones. We would like to stress that in
this context few analytical results are known
\cite{japos-2,gill-massar,bbm-mixed}. Our results here complement
and extend the analysis carried out by some of the authors
in~\cite{bbm-local}.

\subsection{Fixed Measurements}\label{fixed}
Let us start with the most basic scheme for reconstructing a
qubit:  fixed von Neumann  measurements along 2 orthogonal
directions (say, $x$ and $y$) in the equator of the Bloch sphere
for 2D states, or along 3 orthogonal directions (say, $x$, $y$ and
$z$) for 3D states. This kind of scheme is often called
tomography~\cite{tomography,experiments,qudits}.

Consider $N=2\NN$ ($3\NN$) copies of the state $\ketn$. After
$\NN$ measurements along each one of the directions $\vec{e}_i$,
$i=x,y,(z)$, we obtain a set of outcomes $+1$ and $-1$ with
frequencies $\NN \alpha_i$ and $\NN(1-\alpha_i)$, respectively.
This occurs with probability
\begin{equation}\label{probability-local}
  p_n(\alpha)= \!\!\!\!\!\! \prod_{i=x,y,(z)} \!\!\!
  \left(\begin{array}{c}
                              \NN\\
                              \NN \alpha_i
                       \end{array}
                  \right)\left(\frac{1+n_i}{2}\right)^{\NN\alpha_i}\!\!\!
                  \left(\frac{1-n_i}{2}\right)^{\NN(1-\alpha_i)},
\end{equation}
where  $n_i$ are the projections of the vector $\vn$ along each
direction, $n_i\equiv \vec{n}\cdot \vec{e}_i$, and we have used
the shorthand notation $\alpha=\{\alpha_i\}$. The combinatorial
factor takes into account all the possible orderings of the
outcomes and the remaining factors are the quantum  probabilities,
i.e, the appropriate powers of $\tr[\ketn\bran O(\pm \vec{e}_i)]$.

\begin{figure}
 \psfrag{N}{$N$}
 \psfrag{F}{$F$}
 \includegraphics[width=8cm]{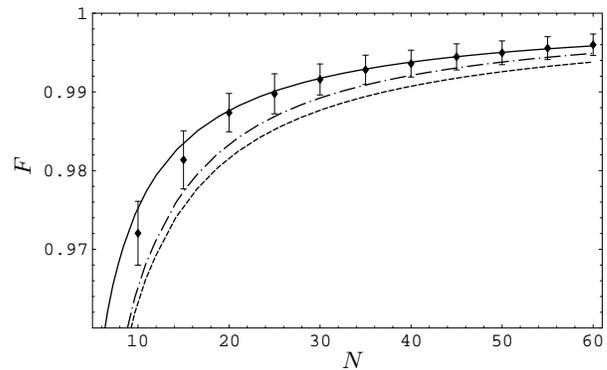}
 \caption {Average fidelities in terms of the number of copies in
 the $2D$ case for
 the optimal collective measurement (solid line), tomographic
 OG (dot-dashed line), tomographic CLG (dashed line) and
 a simulation of the greedy scheme (dots).}\label{fig:2Dfid}
 \end{figure}

Since the expectation value of $\vec{\sigma}$ is
$\bra{\vn}\vec{\sigma}\ketn=\vec{n}$, a straightforward guess
based on the relative frequencies of each outcome is
\begin{equation}\label{cl-guess}
  M^{\rm CLG}_{i}(\alpha)=\frac{
  2\alpha_i-1}{\sqrt{\sum_j (2\alpha_j-1)^2}},
\end{equation}
where the superscript stands for central limit guess (CLG). Notice
the normalization factor which ensures that $|\vM^{\rm CLG}|=1$,
hence $\vM^{\rm CLG}$ always corresponds to a physical pure
state~\cite{likelihood}. The average fidelity in this case is
given by
\begin{equation}\label{fidelity-tomographic}
F=\frac{1}{2}+ \frac{1}{2}\sum_\alpha
          \int dn \; \vn\cdot
                  \vec{M}^{\rm CLG}(\alpha)\, p_n(\alpha),
\end{equation}
where $p_n(\alpha)$ is defined in~\eqref{probability-local}.
Although the CLG~\eqref{cl-guess} is not the \emph{best} state
that one can infer from the data, it has the nice property that it
can be directly (and easily) obtained from the observed
frequencies without further processing. So, it is interesting to
compare its fidelity with the optimal one, given by \eqref{v-optimal} and
\eqref{f-optimal} (see Fig.~\ref{fig:2Dfid}).
%

The OG for this measurement scheme is
$\vec{M}^{\rm OG}(\alpha)=\vec{V}(\alpha)/|\vec{V}(\alpha)|$,
where
\begin{equation}\label{m-guess}
  \vec{V}(\alpha)=
  \int dn\; \vn \ p_n(\alpha),
\end{equation}
and, from \eqref{f-optimal}, the optimal fidelity reads
$F=[1+\sum_{\alpha} V(\alpha)]/2$. Closed expressions of the
fidelity for the lowest values of $N$ can be derived
from~\eqref{fidelity-tomographic} and~\eqref{m-guess} using
\begin{equation}\label{n-integration}
   \int dn \; n_{i_1}n_{i_2}\cdots n_{i_q}=
   \frac{1}{K_q} \delta_{\{i_1 i_2}\delta_{i_3 i_4}\cdots
   \delta_{i_{q-1}i_q\}},
\end{equation}
where the normalization factor is $K_q=q!!$ in 2D and
$K_q=(q+1)!!$ in 3D, and the indexes in curly brackets are fully symmetrized, e.g.,
$\delta_{\{i_1 i_2}\delta_{i_3 i_4\}}=\delta_{i_1 i_2}\delta_{i_3
i_4}+ \delta_{i_1 i_3}\delta_{i_2 i_4}+\delta_{i_1 i_4}\delta_{i_2
i_3}$. Obviously, the integral~\eqref{n-integration} vanishes for
$q$ odd.  For larger values of $N$ the expressions became rather
involved and we have resorted to a numerical calculation.

The 2D case is illustrated in Fig.~\ref{fig:2Dfid}, where the
average fidelity for the above two guesses and $N$ in the range
$10-60$ is shown. In both instances the fidelity approaches unity
as $N$ increases, and the OG always performs better  than the
CLG, as it should.
Notice that to make the graphs more easily readable we have
interpolated between integer points.

\begin{figure}
\psfrag{N}{$N$} \psfrag{SE}{$\epsilon_N$}
\includegraphics[width=8cm]{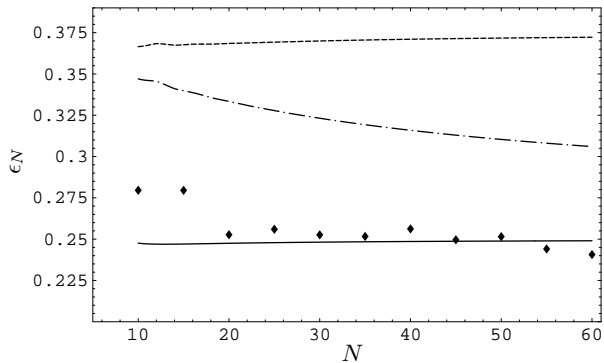}
\caption {Scaled error $\epsilon_N=N(1-F)$ in the 2D case for
collective (solid line), OG (dot-dashed line), CLG (dashed line)
and greedy (diamonds) schemes.}\label{fig:2Derr}
\end{figure}

At this point, it is convenient to define the scaled error
function
\begin{equation}\label{epsilon-N}
\epsilon_N=N(1-F),
\end{equation}
and the limit
\begin{equation}\label{epsilon-limit}
    \epsilon=\lim_{N\to \infty}\epsilon_N ,
\end{equation}
which gives the first order coefficient of the fidelity in the
large $N$ expansion, $F=1-\epsilon/N+ \cdots$
(the asymptotic behavior will be properly discussed in Sec.~\ref{asymptotic}).
Fig.~\ref{fig:2Derr} shows $\epsilon_N$ as a function of $N$ for
2D states.
%
%
 One readily sees that the CLG gives
$\epsilon^{\rm{CLG}}\approx 3/8$~\cite{bbm-local}, while for
collective measurements one has $\epsilon^{\rm{COL}} \approx 1/4$,
in agreement with Eq.~\eqref{fidelity-2D-asymptotics}.  The
stability of the curves $\epsilon^{\rm{COL}}_N$ and
$\epsilon^{\rm{CLG}}_N$ shows that the fidelity is well
approximated by $F=1-\epsilon/N$ for such small values of $N$ as
those in the figure. This asymptotic regime is not yet achieved by
the OG, however we will show in Sec.~\ref{asymptotic} that the OG
gives $\epsilon^{\rm{OG}} = 1/4$, thus matching the collective
bound for large~$N$.

\begin{figure}
\psfrag{N}{$N$} \psfrag{SE}{$\epsilon_N$}
\includegraphics[width=8cm]{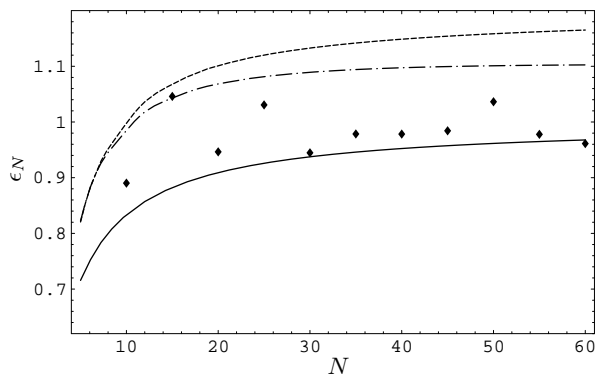}
\caption {Scaled error $\epsilon_N=N(1-F)$ in the 3D case for
collective (solid line), OG (dot-dashed line), CLG (dashed line)
and greedy (diamonds) schemes.}\label{fig:3Derr}
\end{figure}

Fig.~\ref{fig:3Derr} shows the scaled error in the 3D case.
%
%
Again,
one readily sees that the OG performs better than the CLG.
However, the improvement does not seem to be enough to match
the collective bound. We prove analytically in the next section
that $\epsilon^{\rm{OG}}= 13/12> \epsilon^{\rm{COL}}=1$.

In previous paragraphs we have presented the most basic scheme,
i.e., that with a minimal number of orientations of the measuring devices and
without exploiting classical communication. A next step in
complexity is to consider a more general set of fixed directions
$\{m_k\}$. It is intuitively clear that, assuming some sort of
isotropy,  the more directions are taken into account, the better
the estimation procedure will be. For instance, in 2D we may
consider a set of directions  given by the angles $\theta_k=k
\pi/N$, where $k=1,\dots, N$. The set of outcomes $\chi$ can be
expressed as an $N$-digit binary number $\chi=i_{N}i_{N-1}\cdots
i_{2}i_{1}$, where $i_k$ ($=0,1$) and the fidelity then  reads
\begin{equation}\label{isotropic}
    F=\frac{1}{2}+\frac{1}{2}\sum_{\chi=00\cdots0}^{2^{N}-1}\left|\int dn\; \vec{n}
\prod_{k=1}^{N}\frac{1+(-1)^{i_k}  \vn \cdot \vec{m}_k}{2}\right|.
\end{equation}
Analytical results for low $N$ can be  obtained
using~\eqref{n-integration}. For large $N$,  numerical computations
show that this ``isotropic strategy" is indeed better than
tomography (see \cite{monras} for explicit results),
but there is no substantial improvement.

For 3D states, however, it is really important to consider more general
fixed measurement schemes, such as a 3D version of the isotropic one we have  just
discussed.  One can readily see from
Fig.~\ref{fig:3Derr} that the tomographic OG does not saturate
asymptotically the collective bound and one could be tempted to
think that classical communication may be required to attain
it~\cite{bbm-local,gill-massar}. That would
somehow indicate a fundamental difference between the estimation
of 2D and 3D states.

There is a difficulty in implementing the isotropic scheme for 3D states
since the notion of
isotropic distribution of directions is not uniquely defined, which
contrasts with the 2D case. A particularly interesting scheme that
encapsulates this notion (at least asymptotically) and
enables us to perform analytical computations consists of
measurements along a set of random directions. With the same
notation as in \eqref{isotropic}, the fidelity for this set of
directions can be written as
\begin{equation}\label{fid-random}
    F=\frac{1}{2}+\frac{1}{2}\!\!\!\sum_{\chi=0\cdots0}^{2^{N}-1}\int
\prod_{k=1}^{N} dm_k\left|\int
    dn\;
    \vec{n} \frac{1+(-1)^{i_k}  \vn \cdot
\vec{m}_k}{2}\right|.
\end{equation}

\begin{figure}
\psfrag{N}{$N$} \psfrag{E}{$\epsilon_N$}
\includegraphics[width=8cm]{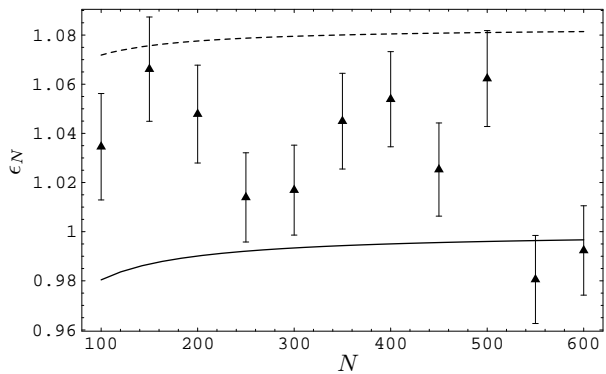}
\caption {Scaled error $\epsilon_N$ of the random scheme
(triangles) as compared to the  OG (dot-dashed line) and the
optimal collective scheme (solid line).}\label{fig:3Dranderr}
\end{figure}

 In Fig.~\ref{fig:3Dranderr} we show the scaled
error, $\epsilon_N$, obtained from numerical simulations for
rather large $N$. One readily sees the improvement of the random
scheme over the tomographic OG. We will  show in the next section that
the former indeed attains the
collective bound asymptotically, thus resolving the puzzle of whether
classical communication is needed or not. A numerical fit gives a
value $\epsilon_N^{\rm{rand}}=1.002\pm 0.008$,  which provides a
numerical check of the analytical results of
section~\ref{asymptotic} below.

\subsection{Adaptive Measurements}\label{adaptive}
In this subsection we will discuss schemes that make use of
classical communication. In principle, they should  be more
efficient than those considered so far.

\subsubsection{One step adaptive}

We first review a method put forward by Gill and
Massar~\cite{gill-massar}, which we call ``one step adaptive".
This scheme, although very simple, suffices to show in a very
straightforward way that local measurements attain the collective
bounds in 2D and 3D. It also has the nice feature that only one
reorientation of the measuring device is required.

The basic idea of the method is to split the measurements in two
stages. In the first one a small number of copies is used to
obtain a rough estimate $\vec{M}_0$  of the state. In the second
stage the remaining copies are used to refine the estimate
by measuring on a plane orthogonal to $\vec{M}_0$. This strategy
has a clear motivation from the information theory point of view.
A measurement can be regarded as a query that one makes to a
system. The most informative queries are those for which the prior
probabilities of each  outcome are the same. Measurements on the
orthogonal plane to $\vec{M}_0$ have this feature. The method
turns out be efficient if the number of copies used in each of the
two stages is carefully chosen.

To be more concrete, suppose we are given $N$ copies of an unknown
qubit state. Let $N_0$ stand for the number of copies used in the
first stage and let $\bar{N}=N-N_0$ stand for the rest. In the 2D
(3D) case, one measures $N_0/2$ ($N_0/3$) copies along two (three)
fixed orthogonal directions and infers the guess $\vec{M}_0$. In
the second stage, one measures $\bar{N}$ ($\bar{N}/2$) along
$\vec{u}$ ($\vec{u},\vec{v}$), which are chosen so that
$\{\vec u,\vec v,\vec M_0\}$ is an orthonormal basis,
and infers the final guess $\vec{M}$.  If
 $N_0\propto N^{a}$ with $0<a<1$, one can show
that the one step adaptive scheme saturates the collective bound
in the asymptotic regime~\cite{gill-massar}. Although we do not
have a rigorous proof, our numerical analysis reveals that the
optimal value of~$a$, i.e., the one that gives the maximal
fidelity, is $a\sim 1/2$. For other choices of~$a$ the scheme can
even be less efficient than some fixed measurement schemes. For
the benefit of the reader, we present a detailed discussion of the
method and a proof of the asymptotic limit within our unified
framework in Appendix~\ref{osa}.

\subsubsection{Greedy scheme}\label{greedy-scheme}
 We now move forward to more  sophisticated schemes  and discuss
 one that exploits much more efficiently classical communication.
 The idea behind it is to
maximize the average fidelity at each single measurement step.
 It is called ``greedy" because  it does not take into
account the total number of available copies, instead, it
treats each copy as if it were the last
one~\cite{fkf,hannemann}.

We first need to introduce some notation. Recall that the set of
outcomes $\chi$ can be expressed as a $N$-digit binary number
$\chi=i_{N}i_{N-1}\cdots i_{2}i_{1}$ ($i_k=0,1$). Since we allow
the $k$-th measurement to depend on the list of previous outcomes, $i_{k-1}
i_{k-2}\cdots i_{2}i_{1}\equiv \chi_{k}$ (note that
$\chi=\chi_N$), we have $\vec{m}(\chi_k)$ instead of $\vec{m}_k$.
This is a compact notation where the length $k$ of the string
$\chi_k$ denotes the number of copies upon with we have already
measured. The orthogonality of the von Neumann measurements is
imposed by the constraint
\begin{equation}\label{von-neumann}
  \vec m(1\chi_{k-1})=-\vec m(0\chi_{k-1}),
\end{equation}
where $1\chi_{k-1}$ is the list of length $k$ obtained by
prepending 1 to the list $\chi_{k-1}$, and similarly for
$0\chi_{k-1}$.
In general, the number of independent vectors for a given $N$ is
$(\sum_{k=1}^{N} 2^k )/2=2^N-1$.  For example, if $N=2$ there are
three independent directions, which can be chosen as
$\vec{m}(0),\vec{m}(00),\vec{m}(01)$,  and the other three are
obtained using Eq.~\eqref{von-neumann}. Since the first
measurement can be chosen at will, this number is  reduced to
$2^N-2$.

The general expression of the conditional probability thus reads
\begin{equation}\label{probability-general}
   p_n(\chi)=\prod_{k=1}^{N}{1+\vec n\cdot\vec m(\chi_{k})\over2},
\end{equation}
and, as discussed in the introduction, the OG gives a
fidelity $F=(1+\Delta)/2$, where
\begin{equation}\label{Delta-greedy}
\Delta=\sum_{\chi=00\cdots0}^{2^{N}-1}\left|\int dn\, \vec{n}\,
p_n(\chi)\right|.
\end{equation}
We could in principle attempt to maximize this expression with
respect to \emph{all} the independent variables, i.e., all
independent $\{\vec{m}(\chi_k)\}$. However, the maximization
process very quickly becomes extremely difficult. In the greedy
scheme one takes a more modest approach: one maximizes at each
step $k$.  This enables us to find a compact algorithm for
computing the fidelity, as we discuss below. Furthermore, we show
in Appendix B that in this situation the optimal local measurement
at each step is indeed of  von Neumann type, i.e., any other POVM
will perform worse.

Let us concentrate on the last step, $N$,  of the greedy scheme.
Suppose we have optimised the previous $N-1$ measurements and
 have obtained a string of outcomes
$\chi_{N-1}$. To ease the notation, let us denote the direction of
the last measurement by $\vec{m}_N$, namely
$\vec{m}_N\equiv\vec{m}(0\chi_{N-1})=- \vec{m}(1\chi_{N-1})$.  We
then need to maximize
\begin{equation}\label{Delta-last}
   d(\chi_N)= | \vec{V}(0\chi_{N-1})|+|\vec{V}(1\chi_{N-1})|.
\end{equation}
Here
\begin{eqnarray}
 \vec{V}(i_N\chi_{N-1})=\int dn\, \vec{n}\, p_n(\chi_{N-1})
   \frac{1+(-1)^{i_N}\vec{n}\cdot \vec{m}_N}{2},
\end{eqnarray}
or, equivalently,
\begin{equation}\label{vectors-last}
   \vec{V}(i_N\chi_{N-1})=
   \frac{1}{2}[\vec{V}(\chi_{N-1})+(-1)^{i_N}\mathsf{A}(\chi_{N-1})\vec{m}_N],
\end{equation}
where $\mathsf{A}$ is the real positive symmetric matrix with
elements
\begin{equation}\label{matrix-A}
    \mathsf{A}_{k l}(\chi_{N-1})=\int dn\, n_k n_l\, p_n(\chi_{N-1}).
\end{equation}
Therefore
\begin{eqnarray}\label{delta-last-2}
    d(\chi_N)&=&\frac{1}{2}\left\{|\vec{V}(\chi_{N-1})\right.+
   \mathsf{A}(\chi_{N-1})\vec{m}_N| + \nonumber \\
     &&
     \left.|\vec{V}(\chi_{N-1})-\mathsf{A}(\chi_{N-1})\vec{m}_N|\right\}.
\end{eqnarray}
Notice that for 2D states and fixed $d(\chi_N)$ the points
$\vec{\mu}=\mathsf{A}\,\vec{m}_N$ lay on an ellipse with focus
at $\pm \vec{V}$ (an ellipsoid for 3D states). In addition
they fulfil the normalization constraint
\begin{equation}\label{omega-constrain}
    \vec{\mu} \cdot(\mathsf{A}^{-2} \vec{\mu})=1,
\end{equation}
which also defines an ellipse (ellipsoid in 3D) centered at the
origin. As usual,  optimality tells us that the maximum of
$d(\chi_N)$ occurs at the points of tangency of the ellipses
(ellipsoids). This provides a geometrical procedure for finding
the optimal direction $\vec{m}_N$ and  an algorithm for computing
$|\vec{V}(\chi_N)|$.

We now proceed to obtain some explicit expressions for low~$N$.
We discuss only the 3D case, as the 2D case is completely
analogous (numerical results for 2D states are shown in
Fig.~\ref{fig:2Dfid} and Fig.~\ref{fig:2Derr}).

 When we only have one copy of the  state,
$N=1$,  the Bloch vector of the measurement can be chosen in any
direction, say  $\vec{e}_x$, i.e.,
$\vec{m}(0)=-\vec{m}(1)=\vec{e}_x$. The explicit computation of
the vector $\vec{V}$ in~\eqref{v-optimal} gives
\begin{equation}\label{N-1-greedy}
    \vec{V}(\chi_1)=\frac{1}{6}\vec{m}(\chi_1),
\end{equation}
and $F=2/3$, as expected from~\eqref{fidelity-3D} [or
\eqref{Delta-2D-max}].

The first non-trivial case is $N=2$. The matrix
$\mathsf{A}(\chi_1)$ reads
\begin{equation}\label{A-2}
    \mathsf{A}_{k l}(\chi_1)=\frac{1}{6} \delta_{k l} \ \ \ \  \chi_1=0,1,
\end{equation}
i.e., $\mathsf{A}(\chi_1)$ is independent of $\chi_1$ and
proportional to the identity.  The maximum of~\eqref{delta-last-2}
occurs for $\vec{m}_2 \perp \vec{e}_x$, so we  choose
$\vec{m}_2=\vec{e}_y$, which means, $\vec{m}(00)=
\vec{m}(01)=\vec{e}_y$, (notice that in general these two vectors
do not need to be equal, they are only required to be orthogonal
to $\vec{m}(0)$]. Because of~\eqref{von-neumann}, we also have
$\vec{m}(10)= \vec{m}(11)=-\vec{e}_y$. The OG reads
\begin{equation}\label{guess-N=2}
 \vec{
 M}^{(2)}(\chi)=\frac{\vec{m}(\chi_2)+\vec{m}(\chi_1)}{\sqrt{2}} ,
\end{equation}
e.g.,  $\vec{M}^{(2)}(01)=[\vec{m}(01)+\vec{m}(1)]/\sqrt{2}
=[\vec{m}(01)-\vec{m}(0)]/\sqrt{2}=[\vec{e}_y-\vec{e}_x]/\sqrt{2}$.
One obtains
\begin{equation}\label{V-modul-N=2}
   |\vec{V}(\chi_2)|=\frac{\sqrt{2}}{12}
\end{equation}
for all~$\chi_2$,
which implies
\begin{equation}\label{fidelity-N=2}
    F^{(2)}=\frac{3+\sqrt{2}}{6}.
\end{equation}

The case $N=3$ can be computed along the same lines. One can
easily see that $\vec{m}(\chi_3)$ has to be perpendicular to
$\vec{m}(\chi_2)$ and $\vec{m}(\chi_1)$. This shows that, up to
$N=3,$ the greedy approach does not use classical communication,
i.e,  the directions of the measuring devices are only required to be
mutually orthogonal, independently of the outcomes. The optimal
guess is a straightforward generalization of \eqref{guess-N=2}:
\begin{equation}\label{guess-N=3}
    \vec
M^{(3)}(\chi)=\frac{\vec{m}(\chi_3)+\vec{m}(\chi_2)+\vec{m}(\chi_1)}{\sqrt{3}},
\end{equation}
and the fidelity reads
\begin{equation}\label{fidelity-N=3}
    F^{(3)}=\frac{3+\sqrt{3}}{6}.
\end{equation}

The above  results could have been anticipated. As
already mentioned, the outcomes of a measurement on the
plane orthogonal to the guess have roughly the same
probability and are, hence, most informative. One can
regard these measurements as corresponding to mutually
unbiased observables, i.e., those for which the overlap
between states of different basis (related to each
observable) is constant~\cite{unbiased}. Hence, there is
no redundancy in the information  about the state
acquired from the different observables. This point of
view also allows to extend the notion of (Bloch vector)
orthogonality to states in spaces of arbitrary dimension.

The case $N=4$ is even more interesting, since four mutually
orthogonal vectors cannot fit onto the Bloch sphere. We expect
classical communication to start playing a role here. Indeed the
Bloch vectors $\vec{m}(\chi_4)$ do depend on the outcomes of
previous measurements. They can be compactly written
as~\footnote{The generalization
of this formula to arbitrary $N$, i.e., $\vec{m}(\chi_N)=(-1)^{i_N}
\sum_{k=1}^{N-2}\vec{m}(\chi_k)\times
\vec{m}(\chi_{N-1})/\sqrt{N-2}$,  provides a set of
points that are roughly isotropically distributed in the sphere,
which may be interesting in other contexts.}
\begin{equation}\label{ebc-vectors}
\vec{m}(\chi_4)=
\frac{(-1)^{i_4}}{\sqrt{2}}\sum_{k=1}^{2}\vec{m}(\chi_k)\times
\vec{m}(\chi_{3}).
\end{equation}
Again, one can see
that the vectors $\vec{m}(\chi_4)$ are orthogonal to the guess one
would have made with the first three measurements. The fidelity in
this case is
\begin{equation}\label{fidelity-N=4}
F^{(4)}={15+\sqrt{91}\over30}.
\end{equation}

For larger $N$, we have computed the fidelity of the greedy scheme
by  numerical simulations. In Fig~\ref{fig:3Derr}
(Fig.~\ref{fig:2Dfid} and Fig.~\ref{fig:2Derr} for 2D states) we
show the results for $10\leq N\leq 60$ (diamonds). Notice that the
greedy scheme is indeed better than fixed measurement schemes and
approaches the collective bound very fast.

Actually, the greedy scheme is the best we can use if we do
not know a priori the number of copies that will be available; obviously,
the best one can do in these circumstances is to optimise at each step. However, if $N$
is known, we have extra information that some efficient schemes
could exploit to increase the fidelity. We next show that this is
indeed the case.

\subsubsection{General LOCC scheme}
In the most general LOCC scheme one is allowed to optimise over
all the Bloch vectors $\{m(\chi_k)\}$, thus taking into account
the whole history of outcomes. Up to $N=3$ the results are the
same as for the greedy scheme: orthogonal Bloch
vectors for the measurements and no classical communication required.
The results
\eqref{fidelity-N=2} and \eqref{fidelity-N=3} are, therefore, the
largest fidelity that can be attained by any LOCC scheme.

The most interesting features appear at  $N=4$. Here there are 14
independent vectors, which can be grouped into two independent
families of seven vectors. With such a large number of vectors an
analytical calculation is too involved and we have resorted
partially  to a numerical optimization. The solution exhibits some
interesting properties. First, one obtains that
$\vec{m}(\chi_1)\perp \vec{m}(\chi_2)$,  for all $\chi_1$ and $\chi_2$,
as in the $N=2$ and $N=3$ cases. Therefore one can choose
$\vec{m}(\chi_1)=(-1)^{i_1} \vec e_{x}$ and
$\vec{m}(\chi_2)=(-1)^{i_2} \vec e_{y}$. Only for the third and
fourth measurement one really has to take different choices in
accordance to the sequence of the preceding outcomes. The Bloch
vectors of the third measurement can be parametrized by a single
angle $\alpha$ as
\begin{equation}\label{mx3}
   (-1)^{i_3} \vec{m}(\chi_3)=\cos\alpha \, \vec{u}_1(\chi_2)+\sin\alpha \,
\vec{v}_1(\chi_2),
\end{equation}
where
\begin{eqnarray}
  \vec{u}_1(\chi_2) &=& \vec{m}(\chi_1)\times \vec{m}(\chi_2), \nonumber \\
  \vec{v}_1(\chi_2)&=& \vec{u}_1(\chi_2)\times \vec{s}(\chi_2),\nonumber\\
    \vec{s}(\chi_2)&=&{ \vec{m}(\chi_2)+ \vec{m}(\chi_1)\over\sqrt2}  .
\end{eqnarray}
Notice that, rather unexpectedly, $\vec m(\chi_{1})$, $\vec
m(\chi_{2})$ and $\vec m(\chi_{3})$ are not mutually orthogonal.
The optimal value of this angle is $\alpha=0.502$. Although we
cannot give any insight as to why this value is optimal, in
agreement with our intuition one sees that  $\vec m(\chi_{3})\perp
\vec M(\chi_2)$, i.e., the third measurement probes the plane
orthogonal to the Bloch vector one would guess from the first two
outcomes [see Eq.~\eqref{guess-N=2}]. The vectors of the fourth
measurement can be parametrized by two angles, $\beta$ and $\gamma$, as
\begin{equation}\label{mx4}
   (-1)^{i_4} \vec{m}(\chi_4)=\cos\gamma \, \vec{u}_2(\chi_3)+\sin\gamma \,
\vec{v}_2(\chi_3),
\end{equation}
where
\begin{eqnarray}
\vec{u}_2(\chi_3)&=& \vec{s}(\chi_2)\times \vec{m}(\chi_3),\nonumber \\
   \vec{v}_2(\chi_3)&=& \cos\beta \, \vec{m}(\chi_3)-\sin\beta\,
  \vec{s}(\chi_2)
\end{eqnarray}
The optimal values of these angles are  $\beta=0.584$,
$\gamma=0.538$, 
and the corresponding fidelity is $F^{(4)}_{\rm general}=0.8206$.
This is just $1.5\%$ lower than the absolute bound $5/6=0.8333$
attained with collective measurement, Eq.~\eqref{fidelity-3D}.
Note that this value is slightly larger than the fidelity obtained
with the greedy scheme, $F^{(4)}_{\mbox{\scriptsize
general}}>F^{(4)}_{\mbox{\scriptsize
greedy}}=(15+\sqrt{91})/30\approx 0.8180$.  The extra information
consisting of the number of available copies has indeed been used
to attain a larger fidelity. We conclude that for $N>3$, it pays
to relax optimality at each step, and  greedy
schemes~\cite{fkf,hannemann} are thus not optimal.  We would like
to remark that if, for some reason, some  copies are lost or
cannot be measured, the most general scheme will not be optimal,
since it has been designed for a specific number of copies. We
have also computed the values of the maximal LOCC fidelities for
$N=5,6$:  $F^{(5)}_{\rm general}=0.8450$ and $F^{(6)}_{\rm
general}=0.8637$. Beyond $N=6$ the small differences between this
and the greedy scheme become negligible.

\section{Local schemes in the asymptotic limit}\label{asymptotic}
The asymptotic expression of the fidelity enables us to compare
different schemes independently of the number of copies. If two
schemes have the same asymptotic fidelity, it is justified to say
that they have the same efficiency, and conversely. Here we will
compute such asymptotic expansions. We will show that, asymptotically, classical
communication is not needed to attain the absolute upper bound
given by the maximum fidelity of the most general collective
measurements. Some of the results presented in this section were
obtained by  two of the authors by explicit computations
in~\cite{bbm-local}. Here we will use a statistical approach that
relate the Fisher information~$\mathrm{I}$~\cite{cover} with the
average fidelity $F$. This approach will greatly simplify our
earlier derivations.

We label the independent state parameters by the symbol $\eta$. This symbol
will refer to the two angles $\theta$, $\phi$ for 3D states:
$\eta\equiv (\theta,\phi)$;  and the polar angle $\theta$ for 2D
states: $\eta\equiv\theta$.

Assume that under a sensible measurement and estimation scheme
(we mean by that a scheme that leads to  a perfect determination
of the state when $N\to \infty$, i.e, $F\stackrel{N\to
\infty}{\longrightarrow} 1$) the estimated state is close to the
signal state, that is, their respective parameters
$\hat\eta(\chi)$ and $\eta$  differ by a small amount. In this
Section, a hat ($\hat{\phantom{a}}$) will always refer to
estimated parameters, the fidelity $f_n(\chi)$, Eq.~\eqref{f},
will be denoted by $f_{\eta}(\hat{\eta})$, and similarly, the
probability $p_n(\chi)$ will be written as $p_\eta(\chi)$. Note that the guessed
parameters $\hat\eta(\chi)$ are based on a particular outcome
$\chi$. This dependence will be implicitly understood when no
confusion arises.

The fidelity can be approximated by the first terms of its series
expansion:
\begin{eqnarray}
     f_{\eta}(\hat\eta)\approx
   1 +
    \frac{1}{2}\left.\frac{\partial^2 f}{\partial
     \hat\eta_i \partial
    \hat\eta_j}\right|_{\hat\eta=\eta}(\hat\eta_i-
    {\eta}_i)(\hat\eta_j-{\eta}_j),
\end{eqnarray}
where we have used that  $f_{\eta}(\eta)=1$ and  $\partial
f_{\eta}/\partial \hat{\eta}|_{\hat{\eta}=\eta}=0$. Averaging over
all possible outcomes, we have
\begin{equation}
    \label{eq:fidelityexpanded}
    F(\eta)\approx1+\frac{1}{2}\tr[\mathsf{H}(\eta)\,\mathsf{V}(\eta)],\\
\end{equation}
where
\begin{equation}\label{f-eta}
    F(\eta)\equiv \sum_{\chi} p_{\eta}(\chi)
    f_{\eta}[\hat{\eta}(\chi)],
\end{equation}
is the ``pointwise" fidelity, $\mathsf{H}(\eta)$
is the Hessian matrix of $f_\eta(\hat\eta)$ at $\hat{\eta}=\eta$, and
$\mathsf{V}(\eta)$ is the covariance matrix, with elements
$\mathsf{V}_{ij}(\eta)=\sum_{\chi}
p_{\eta}(\chi)(\hat\eta_i-{\eta}_i)(\hat\eta_j-{\eta}_j)$.

It is well known that the variance of an unbiased estimator is
bounded by
\begin{equation}\label{eq:variancevsfisher}
   \mathsf{V}(\eta)\geq\frac{1}{\mathrm{I}(\eta)},
\end{equation}
the so called Cram{\'e}r-Rao bound \cite{caves,japos-1,gill-massar}, where
the Fisher information matrix $\mathrm{I}(\eta)$ is defined as
\begin{eqnarray}
    \label{eq:fisherinfo}
    \mathrm{I}_{ij}(\eta)&=&\sum_{\chi}p_{\eta}(\chi)\frac{\partial
    \ln p_{\eta}(\chi)}{\partial \eta_i}
    \frac{\partial \ln p_{\eta}(\chi)}{\partial \eta_j}.
\end{eqnarray}
The conditional probability $p_{\eta}(\chi)$  regarded as a
function of $\eta$ is called the likelihood function
$\mathcal{L}(\eta)=p_{\eta}(\chi)$. It is also well known that the
bound~\eqref{eq:variancevsfisher} is attained by the maximum
likelihood estimator (MLE)~\cite{cramer}, defined as
$\hat{\eta}^{\rm MLE}=\mathrm{argmax} \;\cal{L}(\eta)$. Hence
this bound is tight.

A link between the Fisher information and the fidelity is obtained
by combining \eqref{eq:fidelityexpanded} and
\eqref{eq:variancevsfisher}, and noticing that $\mathsf{H}(\eta)$
is negative definite. We thus have
\begin{equation}
    \label{eq:crbound}
    F(\eta)\leq1+\frac{1}{2}\tr\frac{\mathsf H(\eta)}{\mathrm{I}(\eta)}
\end{equation}
to leading order and for any unbiased estimation scheme.

The Fisher information is additive. This means
that  if $p^{(2)}_{\eta}(\chi,\chi')=p_\eta(\chi)p'_\eta(\chi')$,
which happens when we perform two measurements
[say, $\{O(\chi)\}$ and $\{O'(\chi')\}$]
on two
identical states, the Fisher information of the combined measurement
is simply $\mathrm{I}^{(2)}(\eta)=\mathrm{I}(\eta)+\mathrm{I}'(\eta)$.
In particular, for $N$
{\em identical} measurements, we have
$\mathrm{I}^{(N)}(\eta)=N~\mathrm{I}(\eta)$.

Finally, since  the OG is a better estimator, and it
is asymptotically unbiased, we must have
\begin{equation}\label{eq:ordering}
    F^{\rm{OG}}(\eta)\simeq F^{\rm{MLE}}(\eta)=1+
    \frac{1}{2 N}\tr \frac{\mathsf{H}(\eta)}{\mathrm{I}^(\eta)}
\end{equation}
to leading order, where the fidelities refer to an estimation
scheme consisting of $N$ identical measurements. Below we use
Eq.~\eqref{eq:ordering} to compute the asymptotic limits of the
fixed measurement schemes discussed in Sec.~\ref{fixed}.

We would like to remark that the Cram{\'e}r-Rao bound assumes some
regularity conditions on the figure of merit and the estimators.
These conditions are satisfied for the problem considered here,
but the bound may not hold in more general situations such as the
estimation of a  mixed qubit state (see~\cite{bbm-mixed}).

\subsection{2D states}
The  2D case is rather simple because the states have just one
parameter and the Fisher information is a single number. Moreover,
{\em any} von Neumann measurement whose vector lies on the equator of the Bloch sphere
performed on a 2D system has $\mathrm{I}=1$, as can be checked by
plugging
\begin{equation}
    p_{\theta}(\pm1)=\frac{1\pm\cos(\theta-\theta_m)}{2}
\end{equation}
into \eqref{eq:fisherinfo}, where $\theta_m$ is the polar angle of
$\vec{m}$, the direction along which the von Neumann measurement
is performed.  Therefore, in 2D the Fisher information for a set of $N$
measurements (identical or not) is $\mathrm I^{(N)}=N$.

Since the Hessian is
\begin{equation}\label{eq:hessian2D}
    \mathsf H=\left.\frac{\partial^2
f}{\partial\hat\theta^2}\right|_{\hat\theta=\theta}=-\frac{1}{2},
\end{equation}
{\em any} sensible local measurement scheme on 2D states will yield
\begin{equation}
    F(\theta)=1-\frac{1}{4N}+\cdots .
\end{equation}
Note that  this fidelity is independent of $\theta$, so it
coincides with the average fidelity $F=\int d\theta
F(\theta)/(2\pi) = 1- 4/N$.

We recall that this fidelity is attained by the MLE, and hence
also by the OG. We note that it coincides with the collective
bound, Eq.~\eqref{fidelity-2D-asymptotics}. This  implies  that
the collective bound is attained by tomography,
without classical communication. It is quite surprising that the
most basic scheme, with measurements along two fixed orthogonal
directions, saturates already the collective bound asymptotically.

\subsection{3D states}
The 3D case is more involved. The results shown in
Fig.~\ref{fig:3Derr} hint that tomography does not
saturate the collective bound. There are, however, other
measurement schemes that do saturate this bound and still do not
require classical communication. We prove these two
statements below.
\par
\subsubsection{Tomography}
Tomography, as explained in Sec.~\ref{fixed}, consists
in measuring along three orthogonal directions on the Bloch
sphere. We will concentrate on the OG estimator, or more precisely
on the MLE, which is asymptotically equivalent. We do not discuss
the CLG, Eq.~\eqref{cl-guess}, as it  does not attain the collective
bound even for 2D states (see Fig.~\ref{fig:2Derr}).

We first compute the Fisher information.  Consider a scheme that
consists in repeating $\NN$ times the  following:
take 3 copies of the state and perform a measurement along
$\vec e_x$ on the first copy, along $\vec e_y$ on the second copy,
and along $\vec e_z$ on the third copy (recall that
$N=3\NN$).
These three von Neumann measurements can be regarded as a single measurement  with $2^3$
possible outcomes labeled by $\chi=(\chi_1,\chi_2,\chi_3)$, where $\chi_j=\pm1$. The
probability of obtaining  an outcome $\chi$ is
\begin{equation}
  \label{eq:prob3D}
  p_\eta(\chi)=\prod_{j=x,y,z}\frac{1}{2}\left(1+\chi_j \vec n\cdot e_j\right)   .
\end{equation}
The Fisher information matrix $\mathrm{I}(\theta,\phi)$ of such
elementary measurement is obtained  by substituting
Eq.~(\ref{eq:prob3D}) in Eq.~(\ref{eq:fisherinfo}). Note that the
Fisher information of this scheme is
$\mathrm{I}^{(\NN)}(\theta,\phi)=\NN\, \mathrm{I}(\theta,\phi)$.

The Hessian of the fidelity is
\begin{equation}\label{eq:hessian3D}
    \mathsf{H}(\theta,\phi)=-
    \pmatrix{\displaystyle \frac{1}{2}&0 \cr
     0&\displaystyle\frac{ 1-\cos 2\theta}{4}},
\end{equation}
which turns out to be independent of $\phi$. With this, we obtain
\begin{eqnarray}
    &&\frac{1}{\NN}\tr
        \frac{\mathsf{H}(\theta,\phi)}{\mathrm{I}(\theta,\phi)}=
    \frac{3}{16\NN}  \nonumber \\
    && \times\frac{35+28\cos2\theta+\cos4\theta-8\cos4\phi\sin^4\phi}
    {9+7\cos2\theta-2\cos4\phi\sin^2\theta}.
\end{eqnarray}
Integrating over the isotropic prior probability $dn$, Eq.~\eqref{febc dn}, we obtain
\begin{equation}\label{eq:traceHI}
\frac{1}{\NN}\int dn\, \tr
\frac{\mathsf{H}(\theta,\phi)}{\mathrm{I}(\theta, \phi)}
    =-\frac{13}{18\NN}.
\end{equation}
Recalling (\ref{eq:ordering}) and $N=3\NN$ we  finally get
\begin{equation}
    \label{eq:bound3D}
    F^{\rm OG}=1-\frac{13}{12N}+\cdots.
\end{equation}

As mentioned above, the collective bound $F=1-1/N$ is not attained
by tomography.  At this point, the question arises whether
classical communication is necessary to attain this bound. We next
show that this is not the case.

\subsubsection{Random scheme}
We now consider the so called random scheme i.e, a scheme in
which measurements are performed along random directions chosen
from an isotropic distribution. In contrast to tomography,
which only takes into account three fixed directions, this
scheme explores the Bloch sphere isotropically  if a large number
of copies is available. Therefore one can expect that it will
perform much better.

This approach is equivalent to performing a covariant (continuous)
POVM on each one of the copies separately. Here,  we instead
regard it as  von Neumann measurements and a classical ancilla,
e.g., a ``roulette", that tells us along which direction we
measure. From this point of view the outcome parameters are given
by $\chi=(\xi,(u\equiv\cos\vartheta), \varphi)$, where $\vartheta$
and $\varphi$ are the azimuthal and polar angles of  the direction
$\vec{m}(u,\varphi)$ of the measurement, and $\xi=\pm 1$ is the
corresponding outcome.

Let us compute the Fisher information for a specific state
$\eta=((v\equiv\cos\theta),\phi)$. Since this strategy is
isotropic, the pointwise fidelity $F(\eta)$ is independent of
$\eta$, and we conveniently choose $\eta=((v=0),0)=\underline{0}$. By
the same argument, no average over $\eta$ will be needed:
$F=F(\eta)$. The probability is given by
\begin{equation}
    p_{\eta}(\chi)=
    \frac{1+\xi~\vec n\cdot\vec m (u,\varphi)}{2}  ,
 \end{equation}
and the Fisher information reads
\begin{equation}
    \label{eq:fisherinfo-cont}
    \mathrm{I}_{ij}(\eta)=\sum_{\xi=\pm 1}\int \frac{du\, d\varphi}{4\pi}
  p_{\eta}(\chi) \frac{\partial
    \ln  p_{\eta}(\chi)}{\partial \eta_i}
    \frac{\partial \ln  p_{\eta}(\chi)}{\partial
    \eta_j}.
\end{equation}
The diagonal elements  read
\begin{eqnarray}
    \mathrm{I}_{vv}(\underline{0})\!\!
    &=&\!\!\frac{1}{8\pi}\!\sum_{\xi=\pm1}\!
    \int\! \frac{u^2du\,d\varphi}{1+\xi\sqrt{1-u^2}\cos\varphi}=\frac{1}{2},\\
    \mathrm{I}_{\phi\phi}(\underline{0})
    \!\!&=&\!\!\frac{1}{8\pi}\!\sum_{\xi=\pm1}\!
    \int \!\frac{(1-u^2)\sin^2\varphi\,du\, d\varphi}{1+\xi\sqrt{1-u^2}\cos\varphi}
    =\frac{1}{2}.
\end{eqnarray}
As for the off-diagonal elements, a straightforward calculation
gives
 \begin{equation}\label{eq:offdiagonal}
    \mathrm{I}_{v\phi}(\underline{0})=\mathrm{I}_{\phi v}(\underline{0})=0,
 \end{equation}
as one could expect, since gaining information on $v$ does not
provide information on $\phi$ and viceversa. The Fisher
information matrix thus reads
\begin{equation}
    \mathrm{I}(\underline{0})={1\over2} \pmatrix{1&0 \cr
                 0&1}.
\end{equation}
The Hessian of the fidelity is
\begin{eqnarray}
    \mathsf{H}_{ij}(\underline{0})=\left.\frac{\partial^2 f}
    {\partial \hat{\eta}_i\partial
    \hat{\eta}_j}\right|_{\hat\eta=\underline{0}}=-\frac{\delta_{ij}}{2}.
\end{eqnarray}
Finally, using~\eqref{eq:ordering} we obtain
\begin{equation}\label{eq:fbound-random}
F^{\mathrm{OG}}= 1-\frac{1}{N}+\cdots .
\end{equation}

We conclude that asymptotically classical communication is not
required to saturate the collective bound: a measurement scheme
based on a set of random directions does the job.

\section{Conclusions}\label{conclusions}
We have presented a selfcontained and  detailed study of several
estimation schemes when a number $N$ of identical copies of a
qubit state is available. We have used the fidelity as a figure of
merit and presented a general framework which enables us to treat
collective as well as local measurements on the same footing.

We have considered two interesting situations: that of a
completely unknown qubit state (3D case), and that of a qubit
laying on the equator of the Bloch sphere (2D case). We have
obtained the optimal measurements and maximum fidelities for the
most general collective strategies. These results, although well
known, were scattered in the literature and rederived several
times. Here we have obtained them within a direct and unified
framework. The solution in the 2D case is strikingly simple, and
can be extended to the case of optimal covariant phase estimation.

These collective schemes yield the ultimate fidelity bounds that
can be achieved by any scheme, thus setting a natural scale of
what is a good or a bad estimation. However, they require a joint
measurement on the $N$ copies, which is usually  very difficult,
if not impossible, to be implemented in practice. The main part of
this paper has therefore focused on measurements that can be
implemented in a laboratory with nowadays technology: local von
Neumann measurements.

In the 2D case we have shown that, quite surprisingly, the most
basic tomographic scheme, i.e, measurements along two fixed
orthogonal directions with the adequate data processing (the~OG),
gives already a fidelity that is asymptotically
equal to the collective bound. We have  obtained this limit using
the Cram{\'e}r-Rao bound, which is particularly simple to compute in
this situation. This result is in agreement with the direct (and
lengthier) computation presented in~\cite{bbm-local}.

For the 3D states, tomography, i.e. measurements along three fixed
orthogonal directions, fails to give the asymptotic collective
bound, even with the best data processing.  The main reason of
this failure is that the Bloch sphere is not explored thoroughly.
We have considered an extension that is asymptotically isotropic:
a series von Neumann measurements along random directions. We have
proved that this scheme, which does not make use of classical
communication, does saturates the collective bound. Hence, we
conclude that  in the large $N$ limit, an estimation procedure
based on local measurements without classical communication does
perform as well as the most efficient and  sophisticated
collective schemes.

We have also discussed local schemes with classical communication,
i.e, schemes in which the measurements are devised  in such a way
that they take into account previous outcomes.  We have studied in
detail the one step adaptive scheme of Gill and
Massar~\cite{gill-massar}, which has very interesting features:
only two measurement orientations are required, adaptivity is only
used once, and the estimation is made by a CLG,
which can be read off from the frequency of outcomes. The economy
of resources in this scheme may raise doubts about its efficiency.
In Appendix~\ref{osa} we give a simple proof that for large $N$
it indeed attains the collective bounds.

We have also studied strategies that make a more intensive use of
classical communication. In the  greedy scheme  optimization is
performed at each measurement  step \cite{fkf,hannemann}. This
scheme is the best approach one can take if the actual number of
available copies is not known. We have given a geometrical
condition for sequentially finding the optimal measurements and
have  proved that they have to be of von Neumann type (see
App.~\ref{greedy}), i.e. no general local POVM's will perform
better in this context. We have illustrated the performance of the
method with numerical simulations and have shown the behaviour of
the optimal collective scheme is reached for very low values of
$N$. This occurs for $N$ as low as $N=20$ in 2D and slightly
above, $N=45$, in 3D.

In the most general scheme we see that up to $N=3$ ($N=2$ in 2D)
there is no need for classical communication: the optimal
measurements correspond to a set of mutually unbiased observables.
For larger $N$, the knowledge of the actual value of $N$ provides
an extra information that translates into an increase of the
fidelity. From the practical point of view, however,  this
difference is negligible already at the level of a few copies
($N\gtrsim 6$).

Our approach may be extended to other situations. For instance, the
problem of estimating (qubit) mixed states, which is much more involved,
can be tackled  along the lines described
here~\cite{bbm-mixed}. It would also be interesting to consider
qudits and check whether a set of  mutually unbiased observables
provides the optimal local estimation scheme when the number of
copies coincides with the number of independent variables that
parametrize the qudit state.

\section*{Acknowledgments}
It is a pleasure to thank M.~Baig for his collaboration at early
stages of this work. We also thank R.~Gill and M.~Ballester for
useful discussions. We acknowledge financial support from
Spa\-nish Ministry of Science and Technology project
BFM2002-02588, CIRIT project SGR-00185, and QUPRODIS working group
EEC contract IST-2001-38877.

\appendix
\section{}\label{osa}
The simplest LOCC approach is exemplified by the ``one step
adaptive" scheme~\cite{gill-massar}. Measurements are performed
along just two different directions, and CLG is
used~\footnote{Although one can consider more sophisticated estimators, such as MLE
or OG, they will not improve significantly
the estimation.}. The scheme has two stages and classical
communication is used only once, in going from the first stage to
the second. This is therefore, a very economical scheme from the
practical and theoretical point of view. We here review  the
method and give a straightforward and comprehensive proof that it
saturates the collective bound for large $N$. Given the economy of
the scheme, this is not an obvious result at all. We focus  only
on the 3D case, as the simpler 2D case can be  worked out along the
same lines.

 \noindent \textit{First stage}: One performs
$N_0=N^a$ ($0<a<1$) measurements with a sensible estimator, in the
sense of Sec.~\ref{asymptotic}, and obtains an estimation $\vec
M_0$ with a fidelity $F_0$:
\begin{equation}
  \label{fid0}
  F_0=\sum_{\chi_0}\int
  dn \frac{1+\vec n\cdot\vec M_0(\chi_0)}{2} p_n(\chi_0),
\end{equation}
where $\chi_0$ stands for the list of outcomes obtained in this
first stage.

\noindent \textit{Second Stage}: At this point we use the CLG
on the remaining $2\NN
\equiv\bar{N}=N-N_0$ copies by measuring along \emph{two}
perpendiculars directions, $\vec u$ and $\vec v$, on the plane
orthogonal to $\vec M_0$. In this basis the final guess can be
written as
\begin{equation}
    \label{eq:ansatz-osa}
    \vec M=\vec M_0(\chi_0)\,\cos\omega+
   (\vec u \cos\tau+\vec v \sin\tau) \sin\omega.
\end{equation}
This parametrization ensures that $\vec M$ is unitary. The angles
$\omega$ and $\tau$ depend on the outcomes of this second stage,
which are the frequencies $\alpha_i \NN$, $(1-\alpha_i)\NN$,
($i=u,v$). The probabilities are given by $p_n(\alpha)$ in
\eqref{probability-local}, with $n_u=\vec{n}\cdot\vec{u}$ and
$n_v=\vec{n}\cdot\vec{v}$.  As argued above, we measure on the plane orthogonal to
$\vec{M}_0$ because the two outcomes of each measurement have
roughly the same probability, $\alpha_i \approx 1/2$,  and they
are most informative. It is convenient to define the two
dimensional vector $\vec{r}$, with components:
\begin{equation}\label{r-vector}
   r_i \equiv 2 \alpha_i -1, \qquad i=u,v,
\end{equation}
which, on average, is close to $\vec{0}$. This vector gives an
estimation of the projection of the signal Bloch vector $\vec{n}$
on the measurement plane ($uv$ plane). Hence, $\omega$ is expected
to be small ($\vec{M}_0 \approx \vec M $) and we make the ansatz
\begin{equation}
    \label{omega}
    \omega=\lambda\sqrt{r_u^2+r_v^2}, \qquad
    \tan\tau=\frac{r_v}{r_u},
\end{equation}
where the positive parameter $\lambda$ will be determined later.

The final fidelity for a signal state $\vec{n}$ and outcomes
$(\chi_0,\vec{r})$ is
\begin{equation}
f_n(\chi_0,\vec{r})=\frac{1+\vec n\cdot\vec M(\chi_0,\vec{r})}{2},
\end{equation}
and the average fidelity $F$ reads
\begin{equation}
  \label{fidelity}
  F=\sum_{\chi_0,\vec{r}}\int dn\frac{1+\vec n\cdot\vec
  M(\chi_0,\vec{r})}{2}p_n(\chi_0)p_n(\vec{r}|\chi_0).
\end{equation}
Notice that the probability of obtaining the outcome
$\vec r$, namely, $p_n(\vec r|\chi_0)$
[$\equiv p_n(\alpha)$ in~\eqref{probability-local} with $i=u,v$],
is conditioned on $\chi_0$ through the dependence  of the second
stage measurements
on~$\vec{M}_0(\chi_0)$.

 Since we will compute different averages over $\chi_0$,
$\vec{r}=(r_u,r_v)$ and $\vec{n}$, it is convenient to introduce
the following notation:
\begin{eqnarray}
    \left\langle f \right\rangle_0&=&
    \sum_{\chi_0}~f_n(\chi_0,\vec{r})~p_n(\chi_0),\\
    \left\langle f \right\rangle_r&=&
    \sum_{\vec{r}}~f_n(\chi_0,\vec{r})~p_n(\vec{r}|\chi_0), \\
    \left\langle f \right\rangle_n&=&
    \int dn~f_n(\chi_0,\vec{r}),
\end{eqnarray}
and similarly for averages of other functions of $\chi_0$, $\vec r$ and~$\vec n$.
We will denote composite averaging by simply combining subscripts
(i.e. $\left\langle\left\langle F \right\rangle_r\right\rangle_0
\equiv\left\langle F\right\rangle_{r,0}$). Therefore, we write
\begin{equation}
F\equiv\langle f \rangle_{r,0,n}\equiv\langle f\rangle.
\end{equation}
Since $F=(1 +\Delta)/2$, we have
\begin{equation}\label{eq:f-average-osa}
    \Delta=\langle\vec{n}\cdot \vec M  \rangle.
\end{equation}

In the expansions that we  perform below, we
keep only the terms that contribute to the the fidelity up to order
$1/N$. Recalling that $\omega$ is expected to be small, it follows
that
\begin{eqnarray}
    \cos\omega&=&1-\frac{\lambda^2}{2}\left[r_u^2+r_v^2\right], \\
    \sin\omega\cos\tau&=&\lambda r_u ,\\
    \sin\omega\sin\tau&=&\lambda r_v ,
\end{eqnarray}
to leading order. Therefore, the expectation value in
\eqref{eq:f-average-osa} can be written as
\begin{eqnarray}
    \Delta&=&\left\langle\left
    (1-\frac{\lambda^2}{2}\left\langle r_u^2+r_v^2\right
    \rangle_r\right)\vec n\cdot\vec M_0\right\rangle_{0,n}\\
    &+&\lambda \Big\langle \langle
    r_u\rangle_r~n_u+\langle
    r_v\rangle_r~n_v \Big\rangle_{0,n}.
    \label{expval2}
\end{eqnarray}
Since $r_u$, $r_v$ (or equivalently $\alpha_u$, $\alpha_v$) are
binomially distributed, one readily sees that
\begin{eqnarray}
    \langle r_i\rangle_r&=&n_i, \\
    \langle r_i^2\rangle_r&=& n_i^2+\frac{1-n_i^2}{\NN}.
\end{eqnarray}

We further recall that $\vec n$ is unitary and  that $\{\vec M_0,
\vec u,\vec v\}$ is an orthonormal basis, hence
$n_u^2+n_v^2=1-(\vec n\cdot\vec M_0)^2$, and \eqref{expval2} can
be cast as
\begin{eqnarray}
  && \kern-3.5em \Delta=\lambda+\left[1-{\lambda^2\over2}\left(1+\frac{1}{\NN}\right)\right]
  \left\langle\vec n\cdot\vec M_0\right\rangle_{0,n}
     \nonumber \\
   &&\kern-3.5em -\lambda\left\langle(\vec n\cdot\vec
    M_0)^2\right\rangle_{0,n}+ \frac{\lambda^2}{2}\left(1-\frac{1}{\NN}\right)
    \left\langle(\vec n\cdot\vec M_0)^3
    \right\rangle_{0,n}\kern-0.5em.
\label{febc was here}
\end{eqnarray}
To compute the  moments $ \langle(\vec n\cdot\vec M_0)^{q}
   \rangle_{0,n}$, we consider the angle $\delta$ between $\vec{n}$ and
   $\vec{M}_0$, which is also expected to be small. We have
\begin{equation}
   2 F_0-1= \left\langle\vec n\cdot\vec M_0\right\rangle_{0,n}=
    \left\langle\cos\delta\right\rangle_{0,n}
    \simeq 1-\frac{\left\langle\delta^2\right\rangle_{0,n}}{2},
\end{equation}
where we have used \eqref{fid0}. Therefore
\begin{eqnarray}
    \left\langle(\vec n\cdot\vec M_0)^q\right\rangle_{0,n}&=&
    \left\langle\cos^q\delta\right\rangle_{0,n}
    \simeq1-\frac{q}{2}\langle\delta^2\rangle_{0,n}
    \nonumber \\
    &=&1-2q(1-F_0).
\end{eqnarray}
Now we plug this result back into (\ref{febc was here}) to
obtain
\begin{equation}
    F=\langle f\rangle=1-(1-\lambda)^2(1-F_0)-\frac{1-4(1-F_0)}{2\NN}\lambda^2.
\end{equation}
Since the term $(1-\lambda)^2(1-F_0)$ is always positive, the
maximum fidelity is obtained with the  choice  $\lambda=1$,  and
we are left with
\begin{equation}
    F=1-\frac{1-4(1-F_0)}{2\NN}.
\end{equation}
Since the first estimation is asymptotically unbiased,
\mbox{$1-F_0$} vanishes for large $N_0$ (i.e., for large $N$) and
\begin{eqnarray}
    F&\simeq&1-\frac{1}{2\NN}.
\end{eqnarray}
Recalling that $\NN=(N-N^a)/2$, we finally  have
\begin{equation}
    F=1-\frac{1}{N}+\cdots.
\end{equation}
This concludes the proof.

\section{}\label{greedy}
\newcommand{\E}{\mathrm{E}}
In this appendix we prove that in the greedy scheme the optimal
individual measurements on each copy are of  von Neumann type. We
sketch the proof for 2D states. The 3D case can be worked out
along the same lines.

The history of outcomes  will be denoted, as usual,  by~$\chi$.
Notice that here we consider general local measurements (local
POVMs) with $R$ outcomes, where $R$ is possibly larger than two.
Therefore $\chi$ is a $N$-digit integer number in base $R$: $\chi=i_N
i_{N-1}\cdots i_1$ ($i_k=0,1, \ldots , R-1$). As in
Sec.~\ref{adaptive} we use the notation $\chi_k=i_k i_{k-1}\cdots
i_1$. A measurement on the $k$-th copy is defined by a set of
non-negative rank-one operators
$\{O(\chi_k)\}_{i_k=0}^{R-1}=\{O(i_k\chi_{k-1)}\,|\,i_k=0,1,\ldots,R-1\}$, where
\begin{equation}\label{greedy-povm}
    O(\chi_k)=c(\chi_k) [1+\vec m(\chi_k)\cdot\vec \sigma].
\end{equation}
The non-negative constants $c(\chi_k)$ and the vectors
$\vec{m}(\chi_k)$ are subject to the constraints
\begin{eqnarray}
  \label{eq:const1}
  \sum_{i_k=0}^{R-1} c(\chi_k)&=&1, \\
  \label{eq:const2}
  \sum_{i_k=0}^{R-1} c(\chi_k)~\vec m(\chi_k)&=&0 ,\\
  \label{eq:const3}
  |\vec m(\chi_k) |&=&1,
\end{eqnarray}
which ensures that $O(\chi_k)\ge0$ and $\sum_{i_k}O(\chi_k)=\openone$.
Note that we allow
$c(\chi_k)$ to be zero, thus taking into account the possibility
that each local POVM may have a different number of outcomes without
letting $R$ depend on~$k$.

 Assume we have measured all but the last copy and we wish to
 optimise the last measurement.
Recall from Sec.~\ref{introduction} that the  fidelity can be
written as $F=(1+\Delta)/2$, where
\begin{equation}
    \Delta=\sum_\chi|\vec V(\chi)|,  \quad
\vec V(\chi)=\int  dn~\vec n~p_n(\chi).
\end{equation}
To simplify the notation, let us define  $r\equiv i_N$,
$\vec{m}_r=\vec{m}(r\chi_{N-1})$, and $c_r=c(r\chi_{N-1})$.
Then,
   \begin{equation}
 p_n(\chi)=
    p_n(\chi_{N-1})
    \left[c_r( 1+ \vec n\cdot\vec m_{r})
    \right]
\end{equation}
and
\begin{equation}
    \Delta =
    \sum_{\chi_{N-1}}\sum_{r}|\vec V(r\chi_{N-1})|=
    \sum_{\chi_{N-1}}d(\chi_{N-1})  ,
    \label{eq:greedyfid}
\end{equation}
where we have defined $d(\chi_{N-1})$ as
\begin{equation}\label{eq:greedy-dx}
 d(\chi_{N-1})\equiv\sum_{r}c_{r}\left|\int
 dn~\vec n~p_n(\chi_{N-1})\left(1+\vec
n\cdot\vec m_{r}\right)\right|.
\end{equation}
We further write $\vec V\equiv\vec V(\chi_{N-1})$ and define the
symmetric positive matrix
\begin{equation}
 \mathsf{A}_{ij}\equiv \mathsf{A}_{ij}(\chi_{N-1})\equiv \int
 dn~n_in_j~p_n(\chi_{N-1}).
\end{equation}
Eq.~\eqref{eq:greedy-dx}
becomes
\begin{equation}
\label{eq-greedy-dx-2}
   d=\sum_r c_r|\vec V+\mathsf A \vec m_r|
\end{equation}
(Hereafter the dependency on $\chi_{N-1}$
will be implicitly understood to simplify the notation).

Our task is to maximize~\eqref{eq-greedy-dx-2}. Introducing the
Lagrange multipliers $\lambda$, $\vec{\gamma}$, and $\omega_r$,
the function we need to maximize is actually
\begin{equation}\label{eq:L}
  L=d-\lambda\,\Lambda -\vec{\gamma} \cdot\vec\Gamma -\sum_r
  \omega_r \Omega_r     ,
\end{equation}
where the constraints
\begin{eqnarray}
 \Lambda &=& \sum_r c_r -1, \\
   \vec{\Gamma} &=& \sum_r c_r \vec{m}_r,\\
 \Omega_r &=& \frac{\vec{m}_r^2 -1}{2},
\end{eqnarray}
can be read off from \eqref{eq:const1}, \eqref{eq:const2} and
\eqref{eq:const3}. The factor two in the last expression is
introduced for later convenience. Variations with respect to $c_r$
yield
\begin{eqnarray}\label{eq:max-L}
  \frac{\delta L}{\delta c_r}=|\vec V+\mathsf A\vec m_r|-\vec \gamma\cdot\vec
  m_r-\lambda=0.
\end{eqnarray}
Notice that the points  $\vec{m}_r$ that satisfy this equation
define an ellipse, $\mathcal E$, with focus at $-\mathsf A^{-1}\vec{V}$.
Notice also that the parameter $\lambda$ at the maximum is the value of
$d$ in \eqref{eq-greedy-dx-2} (just multiply \eqref{eq:max-L} by
$c_r$ and sum over $r$ taking into account the constraints~$\Lambda=0$
and~$\vec{\Gamma}=0$).

Finally, consider  the variations of $\vec m_r$ in \eqref{eq:L}.
We obtain
\begin{equation}\label{eq:L-mr}
c_r \left(\mathsf{A} \frac{\vec V + \mathsf{A} \vec{m}_r}{|\vec V +
\mathsf{A} \vec{m}_r|} -\vec{\gamma}\right)=\omega_r \vec{m}_r,
\end{equation}
which means that the vector inside the brackets is proportional to
$\vec{m}_r$. Note that condition $\Omega_r=0$  defines a unit
circle, and the orthogonal vector to this curve is $\vec{m}_r$.
So we only need to prove that the orthogonal vector at point
$\vec{m}_r$ of the ellipse $\mathcal E$ defined in
\eqref{eq:max-L}  has precisely this direction ---this follows
straightforwardly taking variations with respect to $\vec{m}_r$ in
\eqref{eq:max-L}. Therefore, the solution is given by the tangency
points of ellipse $\cal E$  and circle $\Omega_r=0$. There are
only two such points,  they are in opposite directions, and all
constraints and maximization equations are satisfied with
$c_{1,2}=1/2$. This proves that the optimal measurements in the
greedy scheme are indeed von Neumann's~\footnote{For the 3D case
we just have to replace  ellipses by ellipsoids and circles by
spheres.}. Notice that this is a stronger statement than it looks:
local measurements with a larger number of outcomes will perform
worse.

\newcommand{\PRL}[3]{Phys.~Rev. Lett.~\textbf{#1}, #2 (#3)}
\newcommand{\PRA}[3]{Phys.~Rev. A~\textbf{#1}, #2 (#3)}
\newcommand{\JPA}[3]{J.~Phys. A~\textbf{#1}, #2 (#3)}
\newcommand{\PLA}[3]{Phys.~Lett. A~\textbf{#1}, #2 (#3)}

\end{document}